\documentclass[10pt,letterpaper]{article}

\usepackage{graphicx}
\usepackage[hypertexnames=false]{hyperref}
\usepackage[dvipsnames]{xcolor}

\usepackage[square, numbers]{natbib}   
\usepackage{bm}
\usepackage{amsfonts}
\usepackage{amsmath}
\usepackage{amssymb, amsthm, bbold, bm}
\usepackage[font={footnotesize}]{caption}
\usepackage{nameref}
\usepackage{placeins}
\usepackage{algorithm}

\let\classAND\AND
\let\AND\relax
\usepackage{algorithmic}

\let\AND\classAND
\AtBeginEnvironment{algorithmic}{\let\AND\algoAND}
\floatname{algorithm}{Algorithm}

\usepackage[switch]{lineno}

\hypersetup{colorlinks,linkcolor={magenta},citecolor={blue},urlcolor={magenta}}

\title{Modeling memory in time-respecting paths on temporal networks}

\author{Silvia Guerrini, Ciro Cattuto, Lorenzo Dall'Amico\textsuperscript{*}\\
{\normalsize ISI Foundation, 10126, Turin, Italy}\\
{\normalsize * Corresponding to: \texttt{lorenzo.dallamico@isi.it}}}

\date{\today}

\newcommand{\ie}{\emph{i.e., }}
\renewcommand{\algorithmiccomment}[1]{\bgroup\hfill\textbf{/*}~#1 \textbf{*/}\egroup}

\begin{document}

\maketitle

\begin{abstract}
Human close-range proximity interactions are the key determinant for spreading processes like knowledge diffusion, norm adoption, and infectious disease transmission. These dynamical processes can be modeled with time-respecting paths on temporal networks. Here, we propose a framework to quantify memory in time-respecting paths and evaluate it on several empirical datasets encoding proximity between humans collected in different settings. Our results show strong memory effects, robust across settings, model parameters, and statistically significant when compared to memoryless null models. We further propose a generative model to create synthetic temporal graphs with memory and use it to show that memory in time-respecting paths decreases the diffusion speed, affecting the dynamics of spreading processes on temporal networks.
\end{abstract}


\section{Introduction}

Human encounters at short distances are a key driver of social interactions between individuals \cite{kroczek2020interpersonal}. They enable diffusive processes such as knowledge diffusion, norms adoption, and infectious disease propagation \cite{salathe2010}. The unfolding of these diffusive processes depends on the evolution of proximity interactions over time, and calls for appropriate modeling using temporal networks \cite{barrat2013}. In these networks, individuals are modeled as \emph{nodes} and their interactions are represented by \emph{temporal edges} formed by node pairs and a time index indicating when the interaction occurred. Temporal interaction networks of humans display recurrent properties, observed in all collection contexts, such as the bursty behavior and high heterogeneity of temporal patterns \cite{Cattuto_2010, Stehl2011, Fournet2014, genois2015data, DallAmico2022EstimatingHC}. The ubiquity of these observations has attracted the efforts of several researchers who proposed generative models of temporal networks capable of reproducing the observed properties \cite{Perra_2012, starnini2013modeling, Ubaldi_2017, hiraoka2020modeling, le2023modeling, sheng2023, masoumi2024simple}.

\medskip

Here, we focus on describing and modeling long memory patterns, another commonly observed feature of empirical social networks \cite{memory_genois, shapememory}. While in the remainder we will only consider memory in diadic interactions, \ie between pairs of nodes, it is worth mentioning that memory effects have also been recently observed in higher-order structures composed of groups of nodes \cite{gallo2024higher}. Memory in temporal networks influences epidemic spreading \cite{Tizzani_2018, Williams_2019}, diffusion speed \cite{scholtes2014, karsai2011}, and the evolution of the network itself \cite{vedran2016}. Recent works have also shown that accounting for memory leads to improved performance in machine learning tasks such as change-point detection \cite{peixoto2017}, identification of causal temporal timescales \cite{petrovic2023}, and community detection \cite{Rosvall_2014, perri}. As such, memory is a key ingredient of temporal network modeling.
\medskip

Memory in temporal graphs is not uniquely defined, and different approaches have been explored to quantify it. Some works considered self-reinforcement mechanisms in which the existence of a temporal edge is influenced by its past \cite{timevarying}, or by the past of other edges \cite{williams2022}. Using a self-reinforcement mechanism, Ref.~\cite{memory_genois} showed a relation between memory and the observed high heterogeneity of temporal patterns in human proximity networks. Another approach to modeling memory relies on the identification of significant temporal motifs, \ie of temporal patterns among nodes \cite{Longa2024, girardini2025community}. Similarly to Refs. \cite{petrovic2023, Rosvall_2014, perri}, we define memory by relying on \emph{time respecting paths}, which describe the navigability of temporal networks, \ie how information flows from one node to another. Time respecting paths drive the unfolding of dynamical processes \cite{pan} and have been recently used to define a notion of distance between temporal graphs \cite{dallamico2024embeddingbased}. We quantify memory as the likelihood that a time-respecting path returns to a previously visited node. Our definition of memory allows us to detect memory effects in Markovian processes, such as diffusion, induced by a non-Markovian evolution of the graph topology. We propose two models for time respecting paths: one that only includes memory, the other that also accounts for the presence of a community structure. Long memory effects lead to increased model complexity to account for temporal correlations, and some works \cite{peixoto2017, perri} reduce the complexity by assuming memory intervenes only at the mesoscale, \ie on groups of nodes forming communities. Unlike these approaches, both our models have a single parameter encoding memory and do not need the mesoscale assumption for a reduced model complexity.

\medskip

We evaluate our models on eight empirical temporal networks describing human proximity interactions \cite{Fournet2014, Genois2018, Gemmetto2014, hs2015, Vanhems, ozella2021using, Stehl2011}, collected in various contexts by the \texttt{SocioPatterns} collaboration (\cite{Cattuto_2010}, \href{https://www.sociopatterns.org}{sociopatterns.org}). The results show significant memory effects across all networks, as demonstrated by the comparison with null models. Interestingly, we observe comparable memory effects in similar contexts (like schools and workplaces), suggesting a robust behavior of memory across data collections. We further develop a generative model to create synthetic graphs with memory, and demonstrate that increased memory decreases the diffusion speed. Our results enable the numerical quantification of memory, which is, together with the network density and the community structure, a key property in temporal graph modeling.

\section{Results} 
\label{results}

Temporal graphs model complex systems with time-dependent interactions \cite{temporalnetworks}. They are formed by a set of temporal edges $\mathcal{E}$ with elements $(i, j, t, w)$ that denote interactions between node pairs $i, j \in\mathcal{V}$ at time $t$ with weight $w$. This definition accounts time as a discrete variable with resolution $t_{\rm res}$. All interactions within $t_{\rm res}$ are considered simultaneous, and we account for the frequency of interaction by letting the weight $w$ be the total interaction time between two nodes in the time resolution interval.

\medskip

We model non-backtracking time-respecting paths (TRPs) on temporal graphs. In the static setting, a path $\mathcal{P} = \{v_1, v_2,\dots,v_a\}$ is an ordered sequence of nodes that represents the trajectory of a walker on the graph. A \emph{transition} of the walker from one node $v_x$ to the next, $v_{x+1}$, is possible only if $(v_x, v_{x+1})$ is a graph edge. Non-backtracking paths exclude self-loops ($v_{x+1} \neq v_x$) and backtracks ($v_{x+1} \neq v_{x-1}$). TRPs extend this concept by incorporating time and respecting the chronological order of interactions. In a TRP, transitions between nodes at time $t$ are possible only if there is a connection between those nodes at time $t$. Unlike paths on static graphs, they are non-symmetric: a path from $i$ to $j$ does not imply the existence of a path from $j$ to $i$. The size of a TRP can be defined in two alternative ways: one is the \emph{path length} that equals the number of nodes in the path; the other is the \emph{duration}, \ie the time elapsed between the last and first steps in the path \cite{pan}. Both definitions are relevant as they describe properties that directly influence the unfolding of dynamical processing on temporal networks, and crucially determine their outcome \cite{sato2019dyane}. For a more detailed definition of TRPs and their implementation, we refer the reader to Section~\ref{timeproc}. 
We remark that, according to our definition, TRPs preserve the ordering in which path transitions occur and do not constrain the actual time elapsed between consecutive steps.

\medskip

We model the likelihood that a TRP returns to a previously visited node within a \emph{memory horizon}. We observe that TRPs on time-resolved human proximity data display statistically significant memory effects and return to already visited nodes with high likelihood, even if TRPs' steps do not depend on their history. This is possible because TRPs are constrained on the graph's temporal edges.
This effect weakens by increasing the aggregation values $t_{\rm res}$, which impose milder constraints on the path dynamics because the number of edges at each time step can only increase with $t_{\rm res}$. In the \nameref{sec:appendix}, we evaluate the robustness of our results across different resolution values $t_{\rm res}$ and we expand our analysis to a memory definition based on the TRP duration instead of length, which also provides statistically significant results.

\subsection{Modeling memory}
\label{sec:main}

\begin{figure}[!t]
	\centering
	\includegraphics[width=\linewidth]{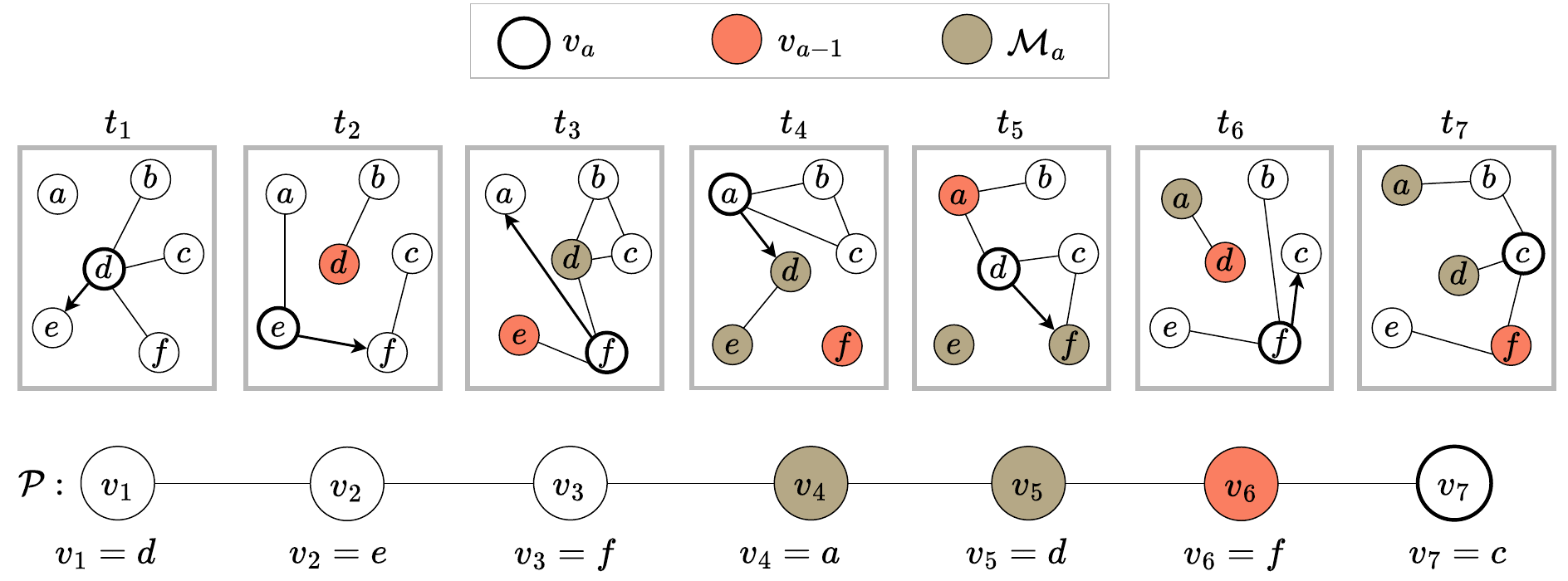}
	\caption{\textbf{Memory in time-respecting paths.} \emph{Top row}. A temporal graph with $n = 6$ nodes. We depict a non-backtracking time-respecting path $\mathcal{P}$ on this graph. The walker is located in the white dot marked with the thick line. An arrow points from this node to the node occupied at the following time step. The red dots indicates the node the path comes from. This node cannot be reached by a path starting from the node marked with the thick line, as it would create a backtrack. The brown dots denote the memory set $\mathcal{M}_a$ at each time step for a memory horizon $m = 4$. Note that the size of the memory set is not constant. \emph{Bottom row}. The path $\mathcal{P}$ at time $t_7$ with the colors and markers obtained at this time.}
	\label{fig:trp_scheme}
\end{figure}

To get our definition of memory, we start from \emph{non-backtracking} TRPs, in which a walker cannot move to the node it came from, \ie $v_{x+1} \neq v_{x-1}$ for all $x$. This choice allows us to focus on long-range memory effect, avoiding short-term back-and-forth transitions between node pairs that may generate fictitious memory effects in the presence of bursty dynamics \cite{saramaki2015exploring}. Considering a path $\mathcal{P} = \{v_1, v_2, \dots, v_a\}$, we let its \emph{memory} $\mathcal{M}_a$ be the set of nodes appearing in the last $m$ path positions, excluding $v_{a}$ and $v_{a-1}$. Formally, given $\mathcal{P} = \{v_1, v_2, \dots, v_a\}$, and $m$, we let $\mathcal{M}_a(\mathcal{P}, a) = \{v_x \in \mathcal{P}~:~2 \leq a-x < m\} = \{v_{a-m+1}, v_{a-m+2},\dots, v_{a-2}\}$. Figure~\ref{fig:trp_scheme} provides a schematic representation of this definition. The parameter $m$ sets the memory horizon in terms of the path length. In the \nameref{sec:appendix}, we explore a definition of $\mathcal{M}_a$ in which the memory horizon is expressed in time units.

\medskip

We propose two models for non-backtracking TRPs. The first (\textbf{MEM}) only accounts for memory and, given $p\in[0,1]$ and $n = |\mathcal{V}|$, the model chooses the next node of $\mathcal{P}$ according to the following rule for $v_{a+1} \notin \{v_a, v_{a-1}\}$:
\begin{equation}
	\label{model1}
	P \left(v_{a+1} |\mathcal{M}_a \right) = p \cdot \frac{\delta \left(v_{a+1} \in \mathcal{M}_a \right) }{\left|\mathcal{M}_a \right|} + \frac{1-p}{n-2}\,\,,
\end{equation} 
and $P(v_{a+1}|\mathcal{M}_a) = 0$ if $v_{a+1} \in \{v_{a}, v_{a-1}\}$. According to this model, with a probability $p$, a node is chosen among those in $\mathcal{M}_a$, and, with probability $1-p$, is randomly sampled from all nodes that can form a non-backtracking path, including those in $\mathcal{M}_a$.

\medskip

Several social networks display a community structure that impacts the statistical properties of TRPs \cite{girvan2002community, saramaki2015exploring, dallamico2024embeddingbased}. We thus propose a generalization of Eq.~\eqref{model1} to account for this effect inspired by the stochastic block model \cite{holland1983stochastic} (\textbf{MEM + SBM}). The stochastic block model generates random graphs with a community structure, by assigning a larger probability of connection between nodes in the same community. Let $\ell: \mathcal{V} \to \{1,\dots,k\}$ be a known labeling function that assigns each node to one of $k$ communities. We let $C \in \mathbb{R}^{k\times k}$ be a symmetric non-negative matrix whose entry $C_{\alpha,\beta}$ describes the affinity between the communities $\alpha$ and $\beta$. Then, according to the stochastic block model, an edge between nodes $u$ and $v$ exists with a probability proportional to $L(u,v) := C_{\ell_u, \ell_v}$. Similarly, the \textbf{MEM+SBM} selects the next node of $\mathcal{P}$ according to the following rule for $v_{a+1} \neq v_a, v_{a-1}$:
\begin{equation}
	\label{model2}
	P \left(v_{a+1} |\mathcal{M}_a \right) = p \cdot \frac{\delta \left(v_{a+1} \in \mathcal{M}_a \right)}{\left|\mathcal{M}_a\right|} +  (1-p)\cdot\frac{L(v_a, v_{a+1})}
	{Z_a}\,\,,
\end{equation}
where $Z_a = \sum_{u \in \mathcal{V} \setminus \{v_a, v_{a-1}\}} L(v_a, u)$ is the normalization and $P(v_{a+1}|\mathcal{M}_a) = 0$ if $v_{a+1} \in \{v_{a}, v_{a-1}\}$. We remark that for both Eqs.~(\ref{model1}, \ref{model2}),  $\sum_{u \in \mathcal{V}}P(u|\mathcal{M}_a) = 1$. Eq.~\eqref{model1} is a particular case of Eq.~\eqref{model2}, obtained for $L_{u,v} = 1$ for all $u,v$, \ie in the absence of a community assignment.
\begin{table}[t!]
	\begin{center}
		{\small
			\begin{tabular}{ l | l | r | r | r }
				Name & Description & n & k & Duration \\
				\hline \hline
				\emph{Primary} \cite{Stehl2011} & Students and teachers from a primary school & $242$ & $11$ & $2$ days \\ 
				\hline
				\emph{High School 1} \cite{Fournet2014} & Students and teachers from a high school & $126$ & $4$ & $4$ days \\
				\hline
				\emph{High School 2} \cite{Fournet2014} & Students from a high school & $180$ & $5$ & $7$ days \\ 
				\hline
				\emph{High School 3} \cite{hs2015} & Students from a high school & $327$ & $9$ & $5$ days \\ 
				\hline
				\emph{Conference} \cite{Genois2018} & Participants to a scientific conference & $405$ & / & $2$ days \\ 
				\hline
				\emph{Office} \cite{Genois2018} & Workers in an office  & $232$ & $12$ & $10$ days \\ 
				\hline
				\emph{Hospital} \cite{Vanhems} & Patients and health-care workers in a hospital & $75$ & $4$ & $5$ days \\ 
				\hline
				\emph{Malawi} \cite{ozella2021using} & Individuals in a village in rural Malawi & $86$ & / & $26$ days \\
				\hline
				\hline
			\end{tabular} 
		}
	\end{center}
	\caption{\textbf{Summary description of the \texttt{Sociopatterns} social networks}. The column ``Name'' reports the name used in the text to refer the dataset. ``Description'' provides concise information on the context. The subsequent columns indicate the number of nodes ($n$), the number of communities ($k$), and the temporal span of the dataset.}
	\label{tab:data_table}
\end{table}

\subsection{Inference on empirical data}

We evaluate our models on eight empirical temporal networks collected by the \texttt{SocioPatterns} collaboration (\cite{Cattuto_2010}, \href{http://sociopatterns.org/}{sociopatterns.org}) describing time-resolved proximity interactions between humans, with a temporal resolution $t_{\rm res} = 20~s$.  These datasets were collected in schools \cite{Stehl2011, Fournet2014, hs2015}, an office \cite{Genois2018}, a hospital \cite{Vanhems}, a scientific conference \cite{Genois2018}, and a rural village in Malawi \cite{ozella2021using}. Some datasets provide additional node categorical attributes: in the school datasets, each node is a student in a known school class; in the hospital and office datasets, nodes have a role attribute. We provide a summary description of these datasets in Table~\ref{tab:data_table}.

We first compare the two models to assess whether the higher complexity of the \textbf{MEM + SBM} model leads to a better goodness-of-fit of the empirical data. We then proceed by describing the maximum likelihood estimator of the memory parameter $p$ for different values of the memory horizon $m$. To assess the significance of these results, we compare the inferred memory $p$ with the one obtained on temporal graphs generated from memoryless null models.

\subsubsection{Goodness-of-fit}

For each temporal graph in Table~\ref{tab:data_table}, we run a collection of non-backtracking TRPs, introduced in Section~\ref{results}. We then use a maximum likelihood estimator to infer the probability of the TRP returning to an already visited node and the community affinity matrix. Section~\ref{timeproc} provides the details on the TRP implementation, and Section~\ref{inference_met} the expression of the maximum likelihood estimators for the \textbf{MEM} and \textbf{MEM + SBM} models.
We evaluate the goodness-of-fit using the Bayesian information criterion (BIC) \cite{modelselection}. Small BIC values imply a better fit.

\medskip

\begin{figure}[t!]
	\centering
	\includegraphics[width=0.9\linewidth]{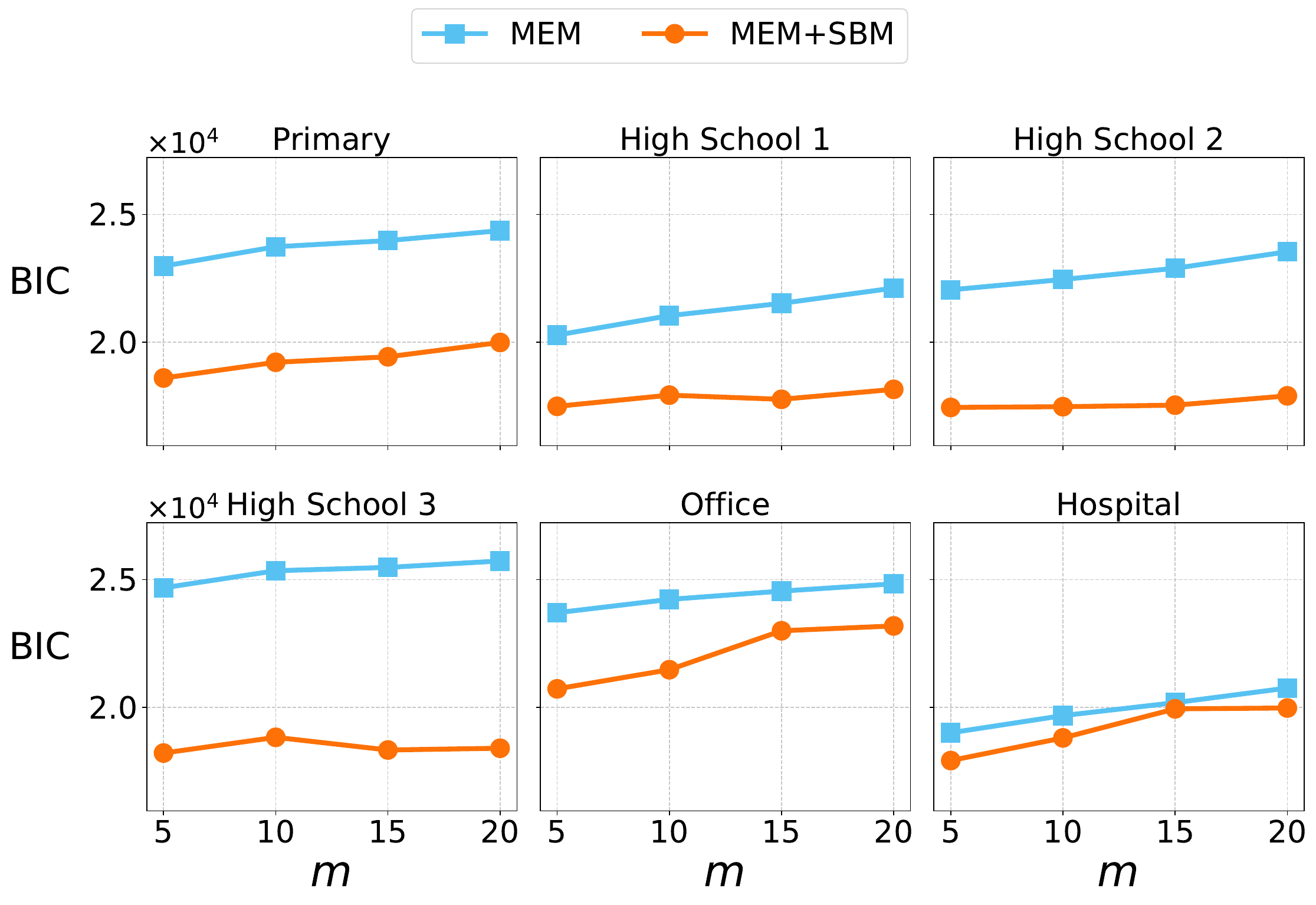}
	\caption{\textbf{Comparison of the goodness-of-fit between \textbf{MEM} and \textbf{MEM + SBM} models as a function of the memory horizon $m$.} Each plot refers to one of the six datasets of Table~\ref{tab:data_table} with a known node label assignment. These dataset have a known node partition in communities. In \emph{Primary}, \emph{High school 1, 2, 3}, each node is assigned to a school class. In the \emph{Office} and in the \emph{Hospital} datasets labels are a role attribute. We show the BIC (lower is better) of the \textbf{MEM} model (cyan squares) and of the \textbf{MEM + SBM} model (orange circles) as a function of the memory length $m$ for $t_{\rm res} = 20~s$. While the plots share the same $y$-axis, the results cannot be compared across datasets.}
	\label{fig:bic_acrossdat20}
\end{figure}

For the six datasets of Table~\ref{tab:data_table} with known node attributes, we compare the BIC for both \textbf{MEM} and \textbf{MEM + SBM} models to evaluate whether the more complex model is required to explain the TRPs statistics. Figure~\ref{fig:bic_acrossdat20} shows the BIC  values for the two models as a function of the memory size, $m$. In all datasets, the \textbf{MEM + SBM} model achieves better results than the \textbf{MEM} model, showing that the community structure needs to be accounted for to explain the TRPs statistic. This is especially evident in the four school datasets that display a highly assortative community structure. Similar results are obtained defining the \emph{memory horizon} in terms of time duration, as reported in the \nameref{sec:appendix}.

\subsubsection{Inference of the memory parameter}
\label{sec:p_inf}

\begin{figure}[t!]
	\centering
	\includegraphics[width=\linewidth]{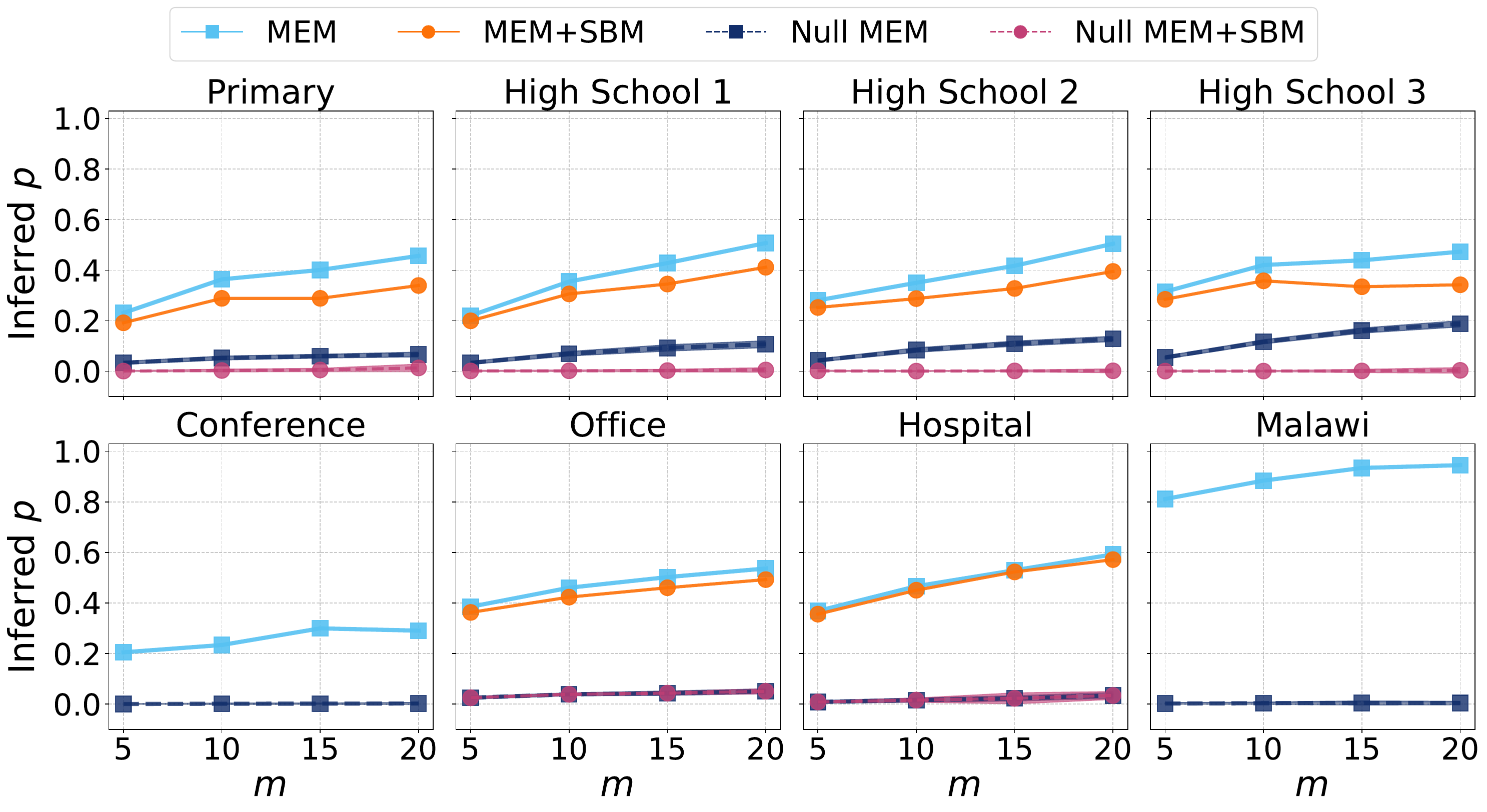}
	\caption{\textbf{Maximum likelihood estimates of the memory  parameter $p$ and comparison with the null models.} Each plot refers to one of the datasets described in Table~\ref{tab:data_table} and shows the inferred value of the probability $p$ of the time-respecting paths to return to a node, as a function of the memory length $m$. The curve ``MEM'' (solid line with squares) is obtained from the \textbf{MEM} model of Eq.~\eqref{model1} on the empirical data. The curve ``MEM+SBM'' (solid line with dots) is obtained from the \textbf{MEM+SBM} model of Eq.~\eqref{model2} on the empirical data. The curve ``Null MEM'' (dashed line with squares) is obtained from the \textbf{MEM} model (Eq.~\eqref{model1}) on the synthetic data, generated from the Erd\H{o}s R\'enyi null model. The curve ``Null MEM + SBM'' (dashed line with dots) is obtained from the \textbf{MEM+SBM} model Eq.~\eqref{model2} on the synthetic data, generated from the SBM null model. For the \emph{Conference} and \emph{Malawi} datasets, we only consider the null model based on Erd\H{o}s-R\'enyi random graphs, while for all other datasets, we consider the one generated from stochastic block model graphs. For these datasets, the labels are known attributes that indicate school classes (\emph{Primary}, \emph{High school 1, 2, 3}), and role attributes (\emph{Office}, \emph{Hospital}). The curves referring to the null models show the mean over $50$ realizations, and the shaded areas represent the standard deviation. For all graphs, we consider the temporal resolution $t_{\rm res} = 20~s$.}
	\label{fig:p_null_acrossdat}
\end{figure}

Figure~\ref{fig:p_null_acrossdat} shows the inferred memory parameter $p$ as a function of the memory size $m$ across datasets. For all datasets with known node labels, we observe the \textbf{MEM} model provides larger memory values $p$ than those observed in the \textbf{MEM+SBM} model. Accounting for the community structure is thus necessary to disentangle homophily from memory. The largest mismatch between the models is observed in the school datasets, coherently with the results shown in Figure~\ref{fig:bic_acrossdat20}. 

Comparing the results across datasets, we observe similar inferred values of $p$ for all schools and in particular, all high schools. Also, workplaces -- \emph{Office} and \emph{Hospital} -- lead to comparable results, with values of $p$ slightly larger than those observed in schools. The high memory observed in \emph{Malawi} and the low one in \emph{Conference} are understood from the experimental context. \emph{Malawi} describes interactions between family members in an African rural village. These interactions are known to be frequent and prolonged \cite{goeyvaerts2018household}, thus explaining the high memory effects observed. \emph{Conference} describes the interactions among researchers at a scientific conference. This can be explained by several characteristics of scientific conferences, like the presence of large gatherings, the schedule of the conference, or the fact that attendees commonly browse around to meet new people.

\medskip

To assess the statistical significance of our results, we compare them with two null models \cite{Gauvin2022} that do not have memory by construction. In the first null model, we obtain a temporal graph by generating a sequence of snapshots drawn from the Erd\H{o}s-R\'enyi model in which each snapshot has the same density as the corresponding one in the empirical dataset. We obtain $p$ as the maximum likelihood estimator of the \textbf{MEM} model on TRPs obtained from this temporal graph. This temporal graph does not have correlations between its snapshots, and any inferred memory effect can be ascribed to randomness. In the second model, we independently generate graphs with a community structure from the \emph{stochastic block model} and estimate $p$ from the \textbf{MEM + SBM} model. The community structure introduces a correlation between the snapshots due to homophily and not memory. Also in this case, each snapshot has the same graph density as the corresponding empirical dataset.
Coherently with the results discussed above, we observe that in the school datasets, the inferred $p$ on the null model with communities is non-zero, which is caused by a highly homophilic community structure. On the contrary, the $p$ inferred from the Erd\H{o}s R\'enyi model is null in all cases. On all datasets, the inferred memory values are much larger than the values inferred from the null models, evidencing a non-trivial memory effect in empirical pathway data.

\subsection{Memory in the TRPs decreases the diffusion speed}

Several works have shown that memory affects the speed of diffusion on temporal networks \cite{karsai2011, rocha2011simulated, artime2017dynamics, williams2022, Tizzani_2018}. We investigate how memory in time-respecting paths (TRPs) impacts the diffusion speed. To do so, we first introduce a model to generate synthetic graphs with varying memory. This allows us to control memory and span different regimes in a controlled setting. We introduce two parameters $\hat{m} \in \mathbb{N}$, and $\alpha \in [0,1]$, describing memory. The parameter $\hat{m}$, similarly to $m$ in the previous sections, is a memory horizon used to introduce correlation between the last $\hat{m}$ snapshot. The parameter $\alpha$ weights the correlation term: by design, we expect the memory $p$ to be an increasing function of $\alpha$. By letting $D_t, A_t$ be the degree and adjacency matrices at time $t$ respectively, we introduce $L_t = (D_t + I_n)^{-1}(A_t + I_n)$, where $I_n$ denotes the identity matrix of size $n$. The entries of $L_t$ indicate the probability of moving between two nodes at time $t$, including self-edges for every node. Consequently, the entries of $M(t; \hat{m}) = \prod_{t' = t-\hat{m}}^{t-1} L_{t'}$ express the likelihood of two nodes to be connected by a TRP in the last $\hat{m}$ steps.

\begin{figure}[!t]
	\centering
	\includegraphics[width=0.95\columnwidth]{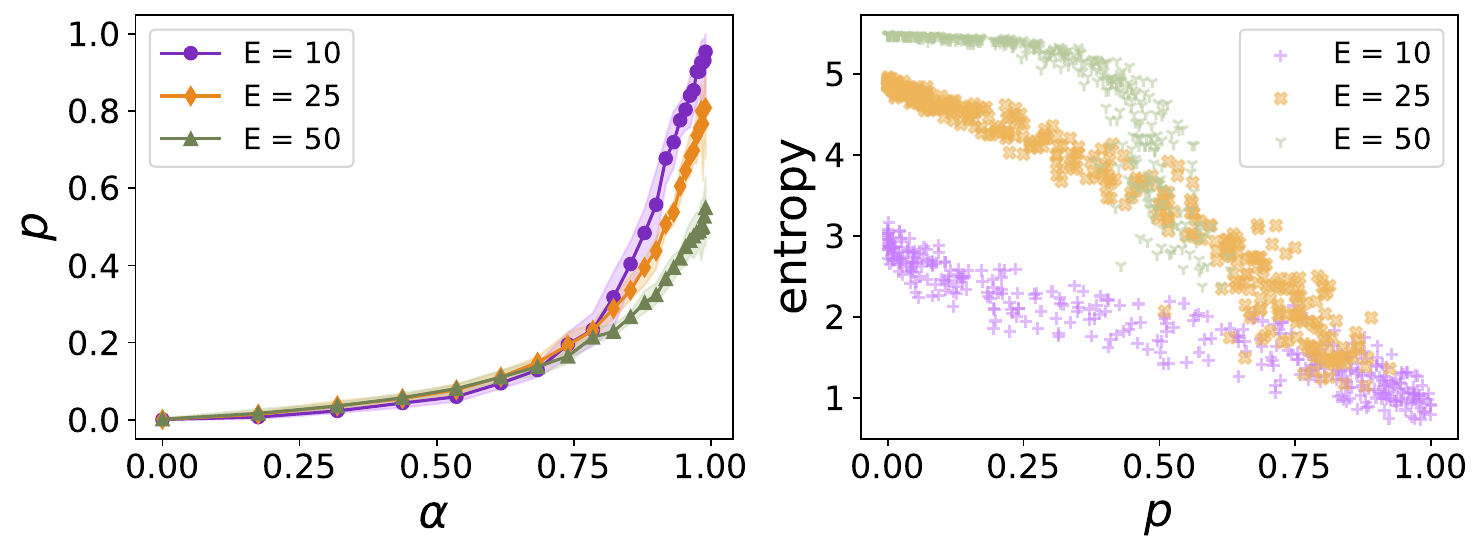}
	\caption{\textbf{The relation between memory and diffusion speed for varying average degree per snapshot, $d$.} \emph{Left panel.} Relation between the weight $\alpha$ appearing in Eq.~\eqref{eq:gen_model} and the estimated $p$ from the \textbf{MEM} model of Eq.~\eqref{model1} with $m = \hat{m} = 5$. The color- and marker-coded lines refer to three values of the graph density, reported in the legend. The solid line is obtained by averaging over $10$ realizations with fixed parameters, while the shaded line is the $5-95$ confidence interval. Each graph has $n = 250$ nodes, and $T = 300$ snapshots. \emph{Right panel.} For each of the graphs of the left panel, we run $15$ simulations of a diffusive process of a signal $\bm{u}$ starting from a randomly selected node at $t = 0$. The figure shows the relation between the inferred $p$ (the same as in the left panel) and the normalized Shannon entropy of the signal density at the end of the process.}
	\label{fig:diffusion}
\end{figure}

We generate the first graph snapshot from the Erd\H{o}s-R\'enyi model and then, for all $t > 1$ and $i\neq j$, the $(i,j)$ entry of the adjacency matrix $A_t$ are set to one independently at random with probability
\begin{align}
	\label{eq:gen_model}
	\mathbb{P}\Big((A_{t})_{i,j} = 1\Big) = d\left[\frac{1-\alpha}{n} + \frac{\alpha}{Z}\Big(M_{ij}(t; \hat{m}) + M_{ji}(t; \hat{m})\Big)\right]\,\,,
\end{align}
where $d$ denotes the average expected degree of each snapshot.
The first summand creates edges between random nodes, while the second enhances the likelihood that two nodes are connected if there is a TRP between them in the last $\hat{m}$ steps. The normalization constant $Z = \frac{1}{n} \sum_{i\neq j}M_{ij}(t;\hat{m})$ balances the two terms. We generate synthetic temporal graphs for varying values of $\alpha$ and infer $p$ using the \textbf{MEM} model introduced in Section~\ref{sec:main}. The left panel of Figure~\ref{fig:diffusion} shows the relation between $p$ and $\alpha$, which is monotonically increasing as expected. We also observe that, while the memory $p$ depends on the weight $\alpha$, it also depends on other graph properties, such as the density. 

\medskip

For each graph, we run $15$ instances of a diffusion process of a scalar quantity on the graph's nodes. We let $u_i$ be the abundance of the diffusing quantity at node $i$ and we let $\bm{u} \in \mathbb{R}^n$ be a vector with entries $u_i$. We consider the following diffusive process:	
\begin{align*}
	\bm{u}_{t+1} = \left(I_n - \beta\left(D_t - A_t\right)\right)\bm{u}_t\,\,,
\end{align*}
where $\beta$ is the diffusion coefficient which we set to $\beta= 0.03$ in the experiments. In each run, we randomly select an active node $s$ at $t = 0$ and let $(u_t)_i = \delta_{is}$. We then compute the normalized Shannon entropy $H(\bm{u}) = -\sum_{i\in\mathcal{V}}u_i~{\rm log}(u_i)$, normalized by ${\rm log}(n)$ as a measure of the diffusion speed. A low entropy means that the final configuration of $\bm{u}$ is close to the initial one, while when the normalized entropy tends to one, it means that the system has reached equilibrium. The right panel of Figure~\ref{fig:diffusion} evidences the determinant role of memory in slowing down diffusion. 

\section{Discussion}

Time-respecting paths (TRPs) drive dynamical processes on temporal graphs \cite{pan}. Here, we focused on high-resolution temporal networks encoding face-to-face proximity between humans, and we introduced a model to quantify memory in TRPs, defined as the probability that a TRP returns to an already visited node. We evaluated our model on several empirical datasets collected by the \texttt{SocioPatterns} collaboration \cite{Cattuto_2010} and inferred memory effects with comparable results across datasets collected in similar settings. We showed that the inferred memory is statistically significant when compared against memoryless null models. TRPs are constrained to the graph topology, which introduces long correlations in the dynamics and generates long memory effects even in Markovian dynamical processes. Memory in TRPs is observed for small aggregation values $t_{\rm res}$. In this case, each graph snapshot tends to be sparse and imposes tight topological constraints on the nodes the TRP can visit. For large aggregations, instead, the snapshot density increases allowing a larger variance in the TRP realization. By introducing a simple model to generate graphs with memory, we explicitly showed that memory in the TRPs slows down the speed of diffusion.

\medskip

Our model provides robust and statistically significant results across datasets, model parameters, and temporal aggregations, but it entails some limitations worth discussing. First, we assumed that a model with time-independent parameters would well describe TRPs, thus that the temporal graph is in a stationary regime. For most of the datasets used in this study -- and in general for most datasets -- this condition cannot be given for granted. For instance, in the school datasets \cite{Stehl2011, Fournet2014, hs2015}, lesson hours are interspersed by breaks with a different dynamics. To refine our results, one can identify change-points in the data \cite{peixoto2017} and add a time-dependent notion of memory. Second, unlike other models \cite{peixoto2017, petrovic2023}, we could describe memory with a low-complexity model, using a single parameter. Its simplicity, however, does not allow the model to capture heterogeneity patterns across nodes that could be included, for instance, by introducing a memory $p$ depending on the node's community class. 

\medskip

We foresee two main directions that follow from our work. One is the design of generative models for temporal networks with memory in the TRPs. The model introduced in Eq.~\eqref{eq:gen_model} takes a first step in this direction, but its objective was primarily to evaluate the relation between diffusion speed and memory in a controlled setting, while a more complex model could generate more realistic synthetic data. Such a model would have applications to better simulate infectious disease spread \cite{Gemmetto2014, cencetti2021digital, colosi2022screening} or to provide an anonymized version of sensitive data encoding proximity between humans \cite{crectu2022interaction, romanini2021privacy}. The second direction is the study and design of accurate aggregation strategies of temporal data. Compared to a time-agnostic framework, modeling dynamical processes at high temporal resolution requires additional efforts in terms of model complexity and data to collect. However, as shown also in this paper, the properties of the temporal network affect the dynamics of dynamical processes unfolding on top of it. Our work calls for the design of efficient procedures to aggregate temporal data that are aware of properties -- such as memory in the TRPs -- that are only defined in the temporal setting.

\section{Methods} \label{sec:methods}

\subsection{Time-Respecting Paths Generative Procedure.} \label{timeproc}

The models proposed in Eqs.~(\ref{model1}, \ref{model2}) describe memory on non-backtracking time-respecting paths (TRPs) on temporal networks. A TRP is a sequence of node-time pairs $\mathcal{P} = \{(v_1, t_1), (v_2, t_2), \dots, (v_a, t_a)\}$ representing the steps of a path on a temporal graph. We require nodes to appear in a chronological order (\ie that $t_{a+1} > t_a$ for all values of $a$) and $(v_a, v_{a+1}, t_a)$ to be a temporal edge for all values of $a$. On top of these conditions, we detail additional definitions of the TRPs that account for known properties of the empirical data under analysis. 

\medskip

First, all datasets of Table~\ref{tab:data_table} span multiple measurement days. We treat each sampling day independently and consider every dataset a collection of temporal graphs. The only exception is the \emph{Hospital} dataset, which we treat as a single temporal graph due to the irregular hours and schedules, with contact occurring almost continuously throughout day and night.
We aggregate time with a resolution $t_{\rm res}$ and consider all events within $t_{\rm res}$ as simultaneous. The edge weight equals the number of edge occurrences within $t_{\rm res}$ before the aggregation, ranging from $0$ to the number of timesteps within the resolution. To initialize the TRP, we first randomly sample an active time $t_1$, \ie a snapshot with at least one temporal edge, and then randomly sample an active node $v_1$, \ie a node with at least one neighbor at $t_1$.

\begin{algorithm}[!t]
	\caption{\texttt{TRP generation}}
	\label{alg:TRP}
	\begin{algorithmic}
		\STATE {\bfseries Input:} temporal graph $\{A_t\}_{t = 1,\dots, T}$; path-length $a$
		\STATE {\bfseries Output:} a TRP $\mathcal{P}$
		\STATE $t_1 \leftarrow$ randomly sample a time with at least one interaction
		\STATE $v_1 \leftarrow$ sample a node with at least one neighbor at $t_1$
		\STATE $v_2 \leftarrow$ sample $v_2$ w.p. $ \propto A_{v_1, v_2}$
		\STATE $\mathcal{P} \leftarrow \{v_1, v_2\}$, initialize the path
		\FOR{$2 \leq x \leq a$}
		\STATE $t_x \leftarrow$ end of the interaction $(v_{x-1}, v_{x})$ 
		\IF{$\mathcal{N}_{v_x}(t_x) \setminus \{v_{x-1}\} = \emptyset$}
		\STATE $t_x \leftarrow \underset{t > t_x}{\rm min} \left|\mathcal{N}_{v_x}(t)\right| >0$
		\ENDIF
		\STATE  $v_{x+1} \leftarrow$ sample $v_{x+1}$ from $\mathcal{N}_{v_x}(t_x)$ w.p. $\propto A_{v_x, v_{x+1}}$
		\STATE $\mathcal{P} \leftarrow$ add $v_{x+1}$
		\ENDFOR
	\end{algorithmic}
\end{algorithm}
%
%
We define the TRP evolution from the first node to account for the interaction and inter-event distributions, which are typically observed to be fat-tailed in temporal social interaction networks \cite{Cattuto_2010, Isella2011}. The broadness of this distribution implies that the time two nodes spend in contact does not have a typical scale. With a non-negligible probability, we can hence observe prolonged interactions between node pairs. By performing a step in the TRP at each time index, \ie letting $t_{a+1} = t_a + 1$, a long interaction between a node pair remains active for several path steps and can be crossed multiple times, confounding a recurrent interaction with a prolonged one. To counter this effect, we let $t_{a+1}$ be the time at which the interaction between $(v_a, v_{a+1})$ is concluded. Consequently, the TRPs we obtain have varying durations. 

The bursty dynamics typically observed in human activity imply that the time distribution between the end of an interaction and the beginning of the next one is also fat-tailed. Burstiness generates TRPs with sequences of node pairs that alternate, introducing fictitious memory effects \cite{saramaki2015exploring}. We thus focus on non-backtracking paths in which $v_{a+1} \notin \{v_a, v_{a-1}\}$, thus alleviating this issue.

\medskip

Algorithm~\ref{alg:TRP} summarizes the TRP generation strategy. The algorithm takes a sequence of time-stamped weighted adjacency matrices and path length $a$ and returns the TRP $\mathcal{P}$. The matrices $A_t$ have size $n \times n$, and the neighborhood of a node $u$ at time $t$ is denoted with $\mathcal{N}_u(t)$. We recall that a TRP has a varying duration, and the path may reach the largest available time before having the desired length $a$. In that case, we regenerate the TRP.

\subsection{Maximum likelihood estimators} 
\label{inference_met}

In this section, we derive the maximum likelihood estimators for the parameters of the \textbf{MEM} and \textbf{MEM + SBM} models. We consider a collection of $R$ TRPs generated independently following the procedure detailed in Section~\ref{timeproc}. For each TRP, we attempt to predict the next node in the sequence. 
By letting $v_{a;r}$ denote the $a$-th step of the $r$-th path, the likelihood of the realizations of the $a+1$ nodes over the $R$ TRPs for the \textbf{MEM} (Eq.~\eqref{eq:L_MEM}) and \textbf{MEM + SBM} (Eq.~\eqref{eq:L_MEMSBM}) models respectively read
\begin{align}
	\mathcal{L}_{\rm MEM} &= \prod_{r = 1}^R P\left(v_{a+1;r}|\mathcal{M}_{a;r}\right) = \prod_{r = 1}^R\left[\frac{p \cdot\delta \left(v_{a+1;r} \in \mathcal{M}_{a;r} \right)}{|\mathcal{M}_{a;r}|}  + \frac{1-p}{n-2}\right] \label{eq:L_MEM}\\
	\mathcal{L}_{\rm MEM +SBM} &= \prod_{r = 1}^R P\left(v_{a+1;r}|\mathcal{M}_{a;r}\right) = \prod_{r = 1}^R\left[\frac{p \cdot\delta \left(v_{a+1;r} \in \mathcal{M}_{a;r} \right)}{|\mathcal{M}_{a;r}|}  + \frac{(1-p)\cdot L(v_{a;r}, v_{a+1;r})}{Z_{a;r}}\right]\,\,. \label{eq:L_MEMSBM}    
\end{align}
In Eq.~\eqref{eq:L_MEMSBM} we used the notation introduced in Section~\ref{sec:main} and denoted with $L \in \mathbb{R}^{n \times n}$ the symmetric non-negative matrix encoding the community affinity between nodes, \ie $L(u, v) = C_{\ell(u), \ell(v)}$. The normalization constant reads $Z_{a;r} = \sum_{u \in \mathcal{V} \setminus \{v_{a;r}, v_{a-1;r}\}} L_{v_{a;r}, u}$. We derive the maximum likelihood estimators of $p$ and $C$ from Eq.~\eqref{eq:L_MEM}, noting that Eq.~\eqref{eq:L_MEM} is obtained from Eq.~\eqref{eq:L_MEMSBM} by letting $L(u, v) = 1$.

\subsubsection*{Parameter $p$}

We compute the derivative of the total log-likelihood with respect to $p$ and set it to zero. We let $\mathcal{R}_{\rm in}$ denote the set of paths where the last node belongs to the memory set, \ie those satisfying $v_{a+1;r} \in \mathcal{M}_{a;r}$. All other paths form the set $\mathcal{R}_{\rm out}$. 
\begin{align*}
	0 =  \frac{\partial {\rm log}(\mathcal{L}_{\rm MEM + SBM})}{\partial p } = \sum_{r \in \mathcal{R}_{\text{out}}}\frac{1}{p-1} + \sum_{r \in \mathcal{R}_{\text{in}}} \left[ \frac{ Z_{a;r} - |\mathcal{M}_{a;r}| \cdot L(v_{a;r}, v_{a+1;r})}{ p \cdot Z_{a;r} + (1-p)\cdot |\mathcal{M}_{a;r}| \cdot L(v_{a;r}, v_{a+1;r})}
	\right]\,\,.
\end{align*}
From this equation, we obtain the following implicit expression of the maximum likelihood estimator of $p$ for the \textbf{MEM + SBM} model:
\begin{align}
	\label{eq:p_est}
	{p} = 1- \frac{|\mathcal{R}_{\text{out}}|}{\displaystyle \sum_{r \in \mathcal{R}_{\text{in}}} \left[ {p} + \left( \frac{Z_{a;r}}{|\mathcal{M}_{a;r}|\cdot L(v_{a;r}, v_{a+1;r})} -1 \right)^{-1} \right]^{-1}}\,\,.
\end{align}
The expression for the \textbf{MEM} model is obtained from Eq.~\eqref{eq:p_est} by letting $L(v_{a;r}, v_{a+1;r}) = 1$ and $Z_{a;r} = n-2$.

\subsubsection*{Community Matrix $C$} 

Following the same procedure, we obtain the maximum likelihood estimator of the matrix $C$, recalling that $L(u,v) = C_{\ell(u), \ell(v)}$. We denote with $q_{\beta;r} = \sum_{u \in \mathcal{V}\setminus\{v_{a;r}, v_{a-1;r}\}} \delta[\beta, l(u)]$ and obtain
\begin{align*}
	&0 = 
	\frac{\partial {\rm log}(\mathcal{L})}{\partial C_{\alpha, \beta}}  \\
	&= \sum_{r \in \mathcal{R}_{\text{out}}} \delta \left[\alpha, \ell(v_{a;r}) \right] \left(\frac{\delta \left[\beta, \ell(v_{a+1;r}) \right]}{C_{\alpha, \beta}} - \frac{q_{\beta;r}}{Z_{a;r}}\right)+\\
	&\,\hspace{2cm}+ (1-p) \sum_{r \in \mathcal{R}_{\text{in}}} \frac{|\mathcal{M}_{a;r}|\delta[\alpha, \ell(v_{a;r})] \Big( \delta[\beta, \ell(v_{a+1;r})]Z_{a;r} - q_{\beta;r}L(v_{a;r}, v_{a+1;r}) \Big)}{p \cdot Z_{a;r}^2 + (1-p) \cdot L(v_{a;r}, v_{a+1;r}) Z_{a;r} \cdot |\mathcal{M}_{a;r}|}\,\,. 
\end{align*}
Again, we can express the maximum likelihood estimator of $C$ with an implicit expression:
\begin{align} \label{eq:mleC}
	{C}_{\alpha, \beta} = \frac{\displaystyle \sum_{r \in \mathcal{R}_{\text{out}}} \delta \left[\alpha, \ell(v_{a;r}) \right] \delta \left[\beta, \ell(v_{a+1;r}) \right]  }{\displaystyle
		\sum_{r \in \mathcal{R}_{\text{out}}} \frac{\delta[\alpha, \ell(v_{a;r})] q_{\beta;r}}{Z_{a;r}} - \sum_{r \in \mathcal{R}_{\text{in}}} \frac{(1-p)\delta[\alpha, \ell(v_{a;r})] \left(Z_{a;r} - L(v_{a;r}, v_{a+1; r}) q_{\beta;r} \right) \cdot |\mathcal{M}_{a;r}| }{p \cdot Z_{a;r}^2 + (1-p) L(v_{a;r}, v_{a+1; r}) Z_{a;r} \cdot |\mathcal{M}_{a;r}|}
	} \,\,,
\end{align}
where we recall once again the dependence of $L$ on $C$.

\subsubsection*{Resolution of the implicit formulas}

As we have already commented, Eqs.~(\ref{eq:p_est}, \ref{eq:mleC}) provide an implicit solution for the maximum likelihood estimators. We detail the procedure adopted to solve these equations. We initialize the parameters, letting $p = 0.5$ and $C$ be the matrix encoding the total number of interactions between communities over the temporal graph duration. We remark that the matrix $C$ (or equivalently of $L$) is defined up to an unknown multiplicative factor, which is handled by the normalization constant $Z_{a;r}$. With an iterative procedure, we then update $p_{\rm new} = g(C, p)$, where the function $g$ is the right hand-side of Eq.~\eqref{eq:p_est} and $C_{\rm new} = f(C, p)$, where $f$ is the right hand-side of Eq.~\eqref{eq:mleC}, looking for the solutions only in the interval $[0,1]$.

\subsubsection*{Availability of data and materials}

All data used in this analysis are public and were downloaded from the \texttt{SocioPatterns} website \href{http://www.sociopatterns.org/datasets/}{sociopatterns.org/datasets/}. The codes to reproduce the results are publicly available in the following repository  \href{https://github.com/SilviaGuerrini/MemoryModel}{github.com/SilviaGuerrini/MemoryModel} where, for convenience, we also share the data we used.

\subsubsection*{Funding}

The authors acknowledge support from the Lagrange Project of the ISI Foundation, funded by Fondazione CRT.

\subsubsection*{Authors' contributions}

SG performed the theoretical analysis, developed computer code, performed the simulations and wrote the first version of the manuscript. LD conceived the project. LD, and CC supervised the project. All authors interpreted the results, reviewed and approved the final version of the manuscript.

\subsubsection*{Acknowledgements}

The authors thank Vincenzo Perri for fruitful discussions.

\FloatBarrier
\newpage
\appendix

\section{Appendix}
\label{sec:appendix}

\subsection*{Inference results for different temporal resolution values}

\begin{figure}[ht!]
	\centering
	\includegraphics[width=\linewidth]{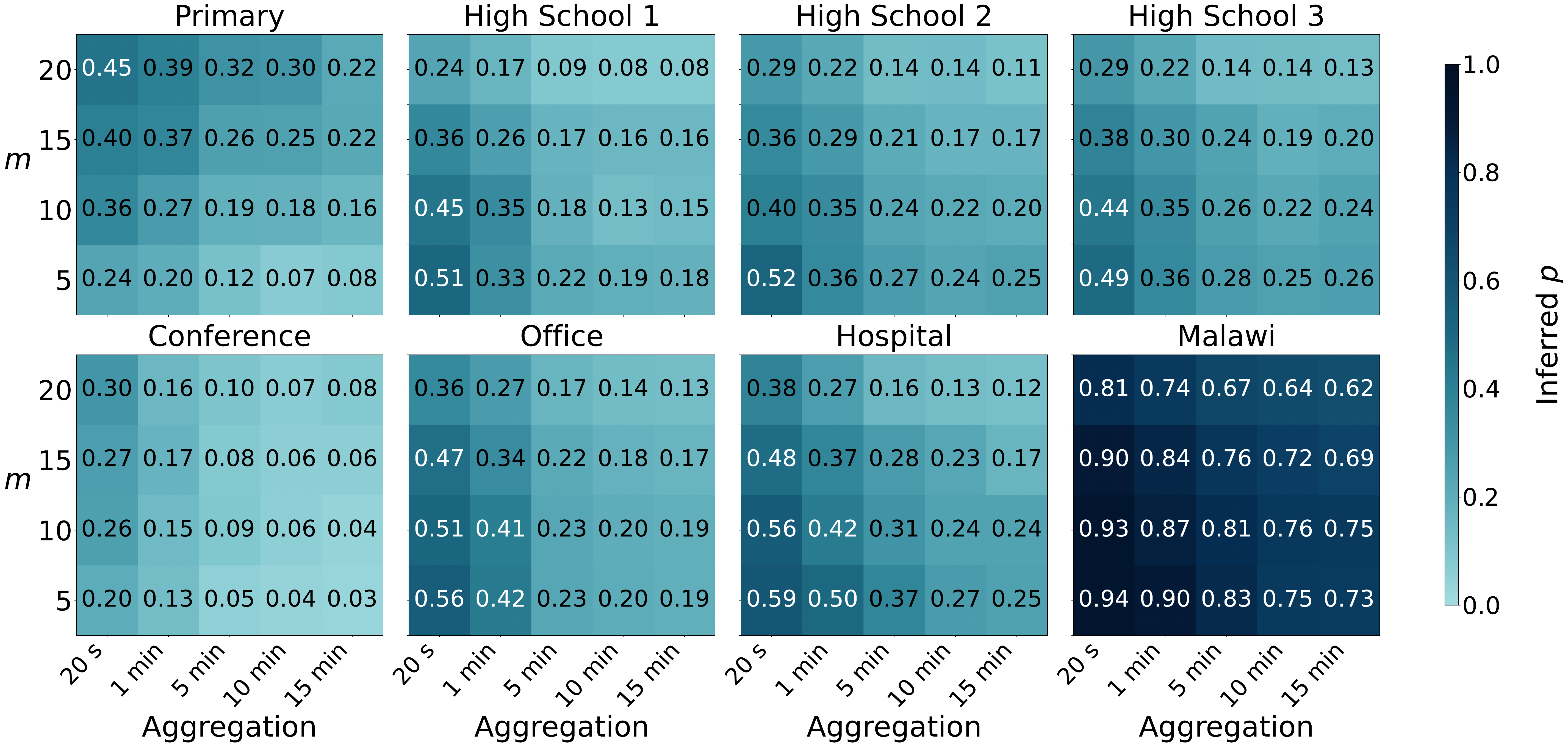}
	\caption{\textbf{Maximum likelihood estimates of the memory parameter $p$.} Each plot refers to one of the datasets described in Table \ref{tab:data_table} and shows the inferred values of the probability $p$ in the \textbf{MEM} model (Eq.~\ref{model1}), probability of the time-respecting paths to return to a node. Paths are generated as per Section~\ref{sec:methods} for different \emph{memory horizon} lengths per path and temporal resolution values $t_{\rm res}$.}
	\label{fig:pmem_across}
\end{figure}

\begin{figure}[ht!]
	\centering
	\includegraphics[width=.9\linewidth]{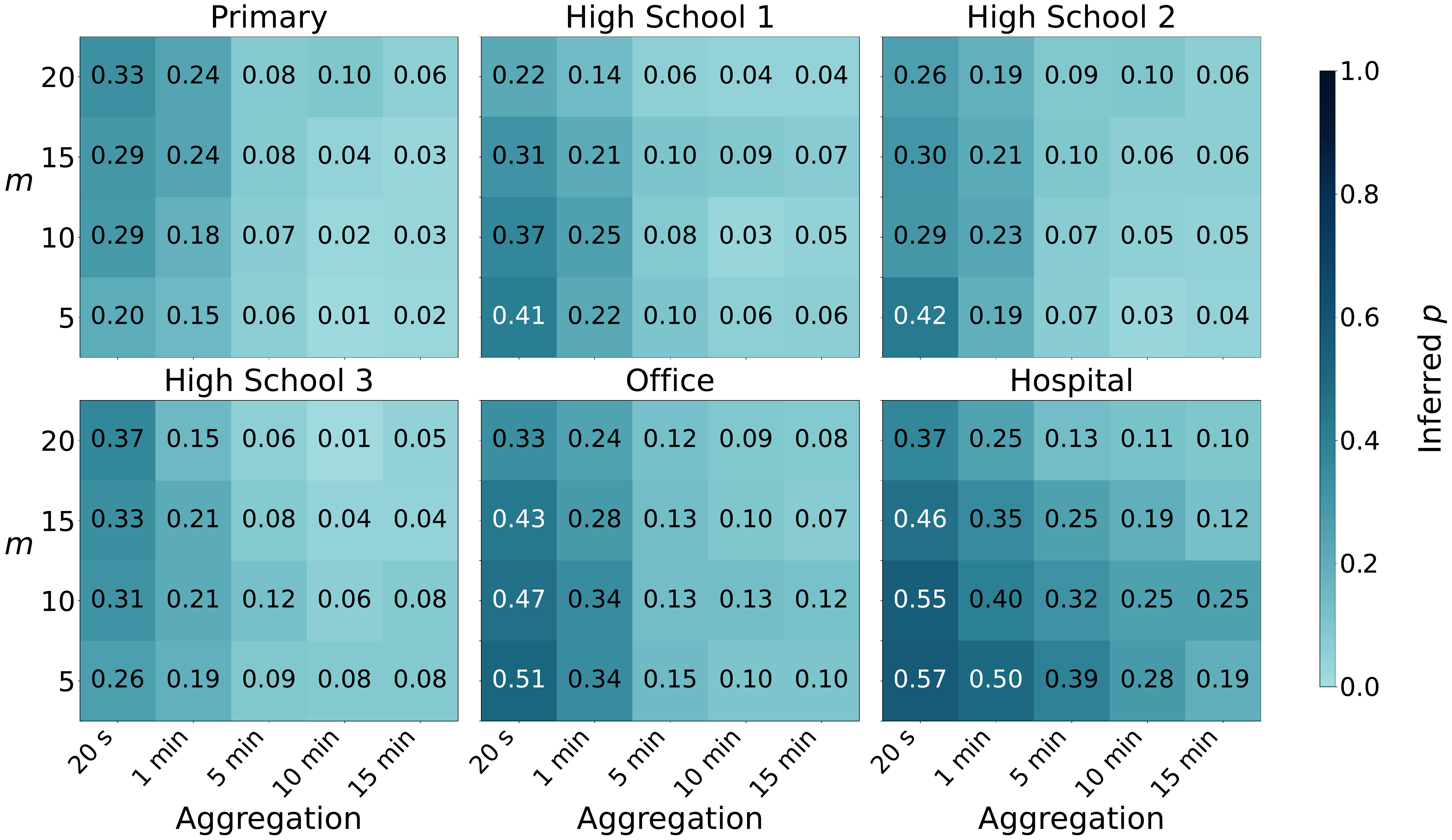}
	\caption{\textbf{Maximum likelihood estimates of the memory parameter $p$.} Each plot refers to one of the datasets described in Table \ref{tab:data_table} and shows the inferred values of the probability $p$ in the \textbf{MEM+SBM} model (Eq.~\ref{model2}), probability of the time-respecting paths to return to a node. Paths are generated as described in Section~\ref{sec:methods} for different \emph{memory horizon} lengths per path and temporal resolution values $t_{\rm res}$.}
	\label{fig:psbm_across}
\end{figure}

TRPs on time-resolved human proximity data exhibit statistically significant memory effects, even if TRPs' steps do not depend on their history. This is possible because TRPs are constrained on the graph's temporal edges. However, this effect deteriorates as the temporal resolution parameter $t_{\rm res}$ increases. Higher values of temporal resolution impose milder constraints on the path dynamics because the number of edges at each time step can only increase with $t_{\rm res}$. In this section, we evaluate the two models across different resolution values $t_{\rm res}$. Figure \ref{fig:pmem_across} shows the maximum likelihood estimates of the memory parameter $p$ in the \textbf{MEM} model (Eq.~\ref{model1}), while figure \ref{fig:psbm_across} in the \textbf{MEM + SBM} model (Eq.~\ref{model2}), across various temporal resolution values. As expected, both figures show a decrease in the inferred memory parameter $p$ as the temporal resolution value $t_{\rm res}$ increases. The effect is already evident between the first two resolution values.

\subsection*{Results for memory expressed in time units}

As mentioned in the main text, the TRP size can be defined in two ways: one is the \emph{path length} that equals the number of nodes in the path; the other is the \emph{duration}, \ie the time elapsed between the last and first steps in the path. In this section, we explore the results for the \emph{memory horizon} expressed in actual time duration (in minutes). 

\begin{figure}[t!]
	\centering
	\includegraphics[width=0.9\linewidth]{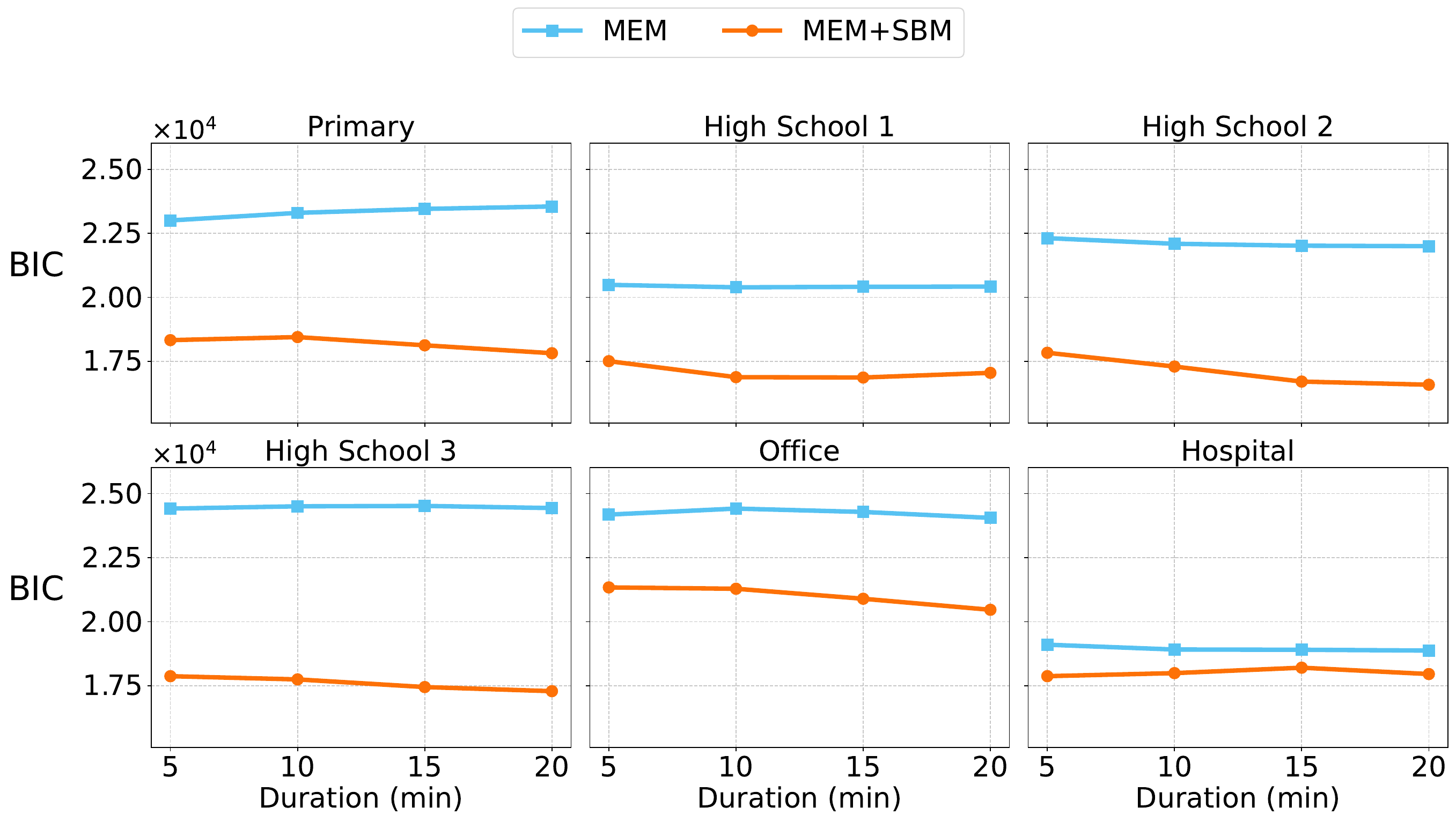}
	\caption{\textbf{Model selection between the \textbf{MEM} and \textbf{MEM + SBM} models.} Each plot refers to one of the six datasets of Table~\ref{tab:data_table} with a known node label assignment. We show the ${\rm BIC}_{\rm MEM}$ (Equation~\eqref{model1}) and ${\rm BIC}_{\rm MEM + SBM}$ (Equation~\eqref{model2}) as a function of the memory time duration (min) for $t_{\rm res} = 20~s$.
		Notice that the scale does not imply comparable results across the datasets.
	}
	\label{fig:bic_acrossdat20_time}
\end{figure}

Figure~\ref{fig:bic_acrossdat20_time} shows the BIC values for the two models as a function of the memory time duration (in minutes). In all datasets the \textbf{MEM + SBM} model achieves better results than the \textbf{MEM} model showing the community structure needs to be accounted for to explain the TRPs statistic. This is especially evident in the four school datasets that display a highly assortative community structure.
\begin{figure}[t!]
	\centering
	\includegraphics[width=0.9\linewidth]{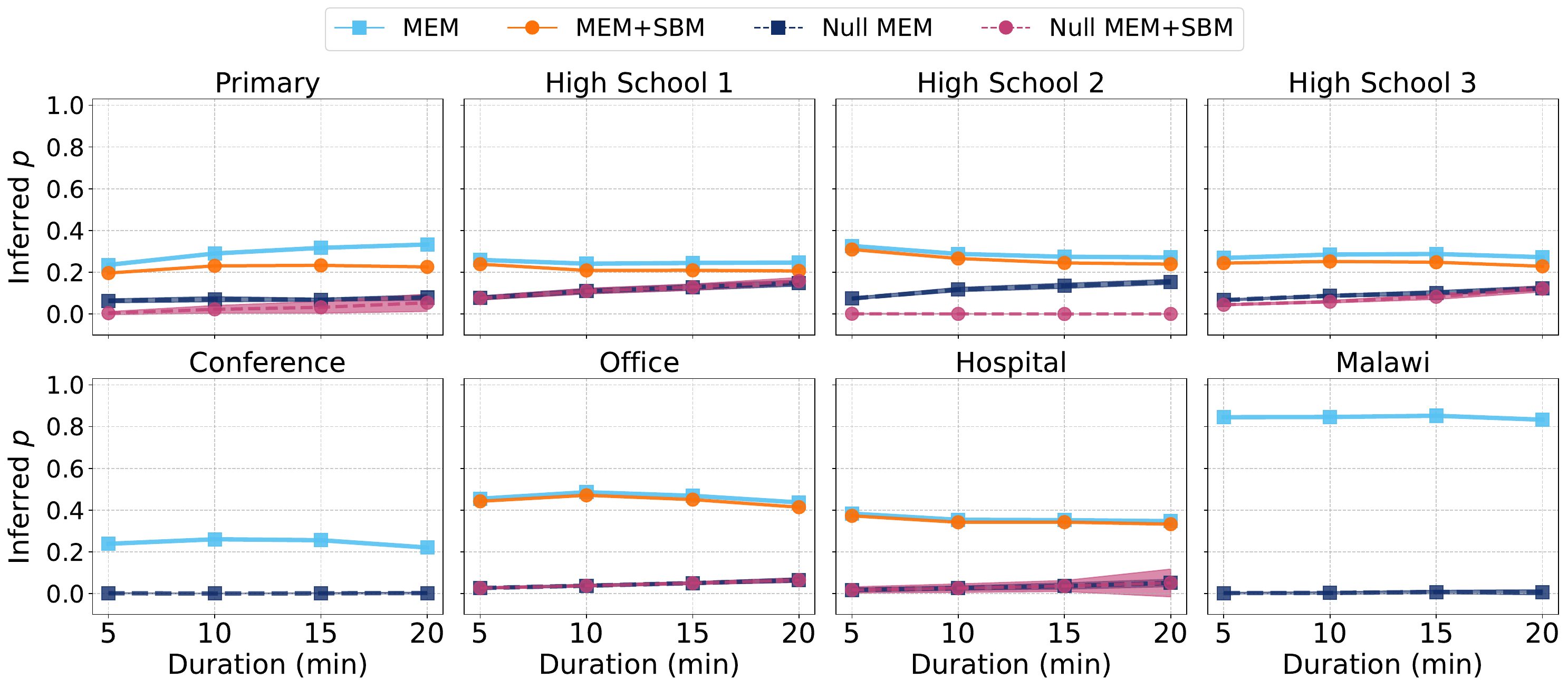}
	\caption{\textbf{Maximum likelihood estimates of the memory  parameter $p$ and comparison with the null models.} Each plot refers to one of the datasets described in Table~\ref{tab:data_table} and shows the inferred value of the probability $p$ of the time-respecting paths to return to a node, as a function of the memory time duration (min). The curve ``MEM'' (solid line with squares) is obtained from the \textbf{MEM} model of Equation~\eqref{model1} on the empirical data. The curve ``MEM+SBM'' (solid line with dots) is obtained from the \textbf{MEM+SBM} model of Equation~\eqref{model2} on the empirical data. For the \emph{Conference} and \emph{Malawi} datasets, we only consider the null model based on Erd\H{o}s-R\'nyi random graphs, while for all other datasets, we consider the one generated from stochastic block model graphs.  
		The curve ``Null MEM'' (dashed line with squares) is obtained from the \textbf{MEM} model (Equation~\eqref{model1}) on the synthetic data. 
		The curve ``Null MEM + SBM'' (dashed line with dots) is obtained from the \textbf{MEM+SBM} model Equation~\eqref{model2}) on the synthetic data. 
		The curves referring to the null models show the mean over $20$ realizations, and the shaded areas represent the standard deviation.
		For all graphs, we consider the temporal resolution $t_{\rm res} = 20~s$.}
	\label{fig:nullmodels_time}
\end{figure}
Figure \ref{fig:nullmodels_time} shows the inferred memory parameter $p$ as a function of the memory time duration across datasets. For all datasets with known node labels, we compare the two models. For the \emph{Conference} and \emph{Malawi} datasets, we report only the inferred memory values from the \textbf{MEM} model. For all schools, \textbf{MEM} provides slightly larger memory values $p$ than those observed in the \textbf{MEM+SBM} model. We observe similar inferred values of $p$ for all schools and in particular all high schools.
All workplaces -- \emph{Office} and \emph{Hospital} -- lead to comparable results, with values of $p$ slightly larger than those observed in schools.
The high memory observed in \emph{Malawi} and the low one in \emph{Conference} is understood from the experimental context. \emph{Malawi} describes interactions between family members in an African rural village. These interactions are known to be frequent and prolonged \cite{goeyvaerts2018household}, thus explaining the high memory effects observed. \emph{Conference} describes the interactions among researchers at a scientific conference. Commonly, attendees browse around to meet new people, naturally decreasing the memory observed in the empirical data.

As we did for the path length configuration, we assess the statistical significance of our results by comparing them with two null models that do not have memory by construction. 
We observe that in all high school datasets, the inferred $p$ on the null models is non-zero, which is probably caused by a highly homophilic community structure. Conversely, in all the other datasets, the memory parameter $p$ inferred from both models is consistently zero.
Across all datasets, the inferred memory values are much larger than the values inferred from the null models, evidencing a non-trivial memory effect in empirical pathway data.
%


\end{document}